\documentclass[10pt,a4paper]{iopart}
\usepackage{iopams}  
\usepackage{bm}
\usepackage{color}
\usepackage[colorlinks=false, pdfstartview=FitV, linkcolor=blue, citecolor=blue, urlcolor=blue]{hyperref}
\def\smallmatrix{\null\,\vcenter\bgroup\vspace@\Let@
 \baselineskip9\ex@\lineskip\ex@
 \ialign\bgroup\hfil$\m@th\scriptstyle{##}$\hfil&&\thickspace\hfil
 $\m@th\scriptstyle{##}$\hfil\crcr}

\begin{document}
\title[]{De Sitter Special Relativity: Effects on Cosmology }
\author{R. Aldrovandi  and J. G. Pereira}

\address{Instituto de F\'{\i}sica Te\'orica, 
S\~ao Paulo State University, Rua Pamplona 145, 01405-900 S\~ao 
Paulo, Brazil}

\begin{abstract} 
The main consequences of de Sitter Special Relativity to the Standard Model of  Physical Cosmology
are examined. The cosmological constant $\Lambda$ appears, in this theory, as a manifestation of  the proper conformal current, which must be added to the usual  energy--momentum density. As that conformal current itself vanishes in absence of sources,   $\Lambda$ is ultimately dependent on the matter content, and can in principle be calculated. A present--day value very close to that given by the crossed supernova/BBR data is obtained through simple and reasonable approximations.  Also a primeval inflation of polynomial type is found, and the usual notion of comoving observer is slightly modified.
\end{abstract}

\section{Introduction} 

Recent years have witnessed a growing interest in  the possibility that  Special Relativity may have to be modified  at ultra--high energies~\cite{dsr}. Reasons have been advanced both from the theoretical and the experimental-observational points of view.  On the theoretical side,  suggestions in that direction come from the physics at the Planck scale, where a fundamental length parameter --- the Planck length $l_P$ --- naturally shows up. Since a length contracts under a Lorentz transformation, the Lorentz symmetry  should be  somehow  broken at that scale~\cite{lorentzX}. On the experimental side, intimations come basically from the propagation of very--high energy photons. More precisely, ultra--high energy extragalactic gamma--ray flares seem to travel slower than lower--energy ones~\cite{magic}. If this comes to be confirmed, it will constitute a clear violation of Special Relativity. 

A well-defined, theoretically self-contained  alternative to usual Special Relativity  is  {\it de Sitter Special Relativity}, in which the 
 Poincar\'e group  is replaced by the de Sitter group~\cite{dssr} as the group supervising kinematics~\cite{nh}.  This leads necessarily to changes in General Relativity, whose basic aspects  have been examined previously~\cite{AP07a,AP07b}.    It has been shown, in particular, that the mentioned problem with gamma--ray flares can be solved. Our objective here is to  reexamine the main implied modifications to the Standard Model of  Physical Cosmology.  
 
Perhaps the main novelty is that on  a general spacetime  $\mathcal{S}$ we have to consider, instead of a tangent  Minkowski space $\mathcal{M}$ at each point $p$, an osculating  de Sitter spacetime  $d\mathcal{S}$ whose pseudo-radius $l$ depends on the energy concentration around $p$.  

The usual source term in Einstein equations is the energy-momentum density tensor $T^{\mu \nu }$, which is the symmetrized Noether current for invariance under translations on $\mathcal{M}$.  Notice that these ``usual'' translations  define transitivity on $\mathcal{M}$. We recall that, in cartesian coordinates $\{x^a\}$ ($a,b,c, ... = 0,1,2,3)$ on $\mathcal{M}$, the Lorentz metric is $\eta_{ab} = $ diag$(1,-1,-1,-1)$ and  the generators of the Poincar\'e group can be written 
\begin{equation}
L_{ab} = \eta_{ac} \, x^c \, P_b - \eta_{bc} \, x^c \, P_a\,; \,\,\, 
L_{a4} = P_a = \partial_a . 
\label{eq:Palgebra}
\end{equation}
The   de Sitter group generators can be realized with the same variables, as
\begin{equation}
L_{ab} = \eta_{ac} \, x^c \, P_b - \eta_{bc} \, x^c \, P_a\,; \,\,\, 
L_{a4} = l P_a - l^{-1} K_a,  
\end{equation}
with
\begin{equation}
K_a = \left(2 \eta_{ac} \, x^c x^b - \sigma^2 \, \delta_a^b \right) \partial_b . 
\label{eq:dSalgebra}
\end{equation}
Quantity ``$l$'' is a natural length parameter~---~the above mentioned  de Sitter  pseudo--radius, or horizon, and $\sigma^2 = \eta_{ab} x^a x^b$.\footnote{\, 
The particular form of (\ref{eq:dSalgebra}) leads to some cases of ``euclideanization":   some expressions exhibit a positive-definite form $\eta^{(+)} = $ diag $(1,1,1,1)$ instead of metric $\eta$. For instance, $K_0 =  2 \, x^0 x^k   \partial_k + \eta^{(+)}_{ab} x^a x^b \partial_0$ ($i, j, k, ... = 1, 2, 3$).  Consequences will appear in the corresponding general--relativistic currents (\ref{eq:confcurrent1}, \ref{eq:concurr}),  in which an euclideanization of the Riemann metric will turn up   [see Eq.(\ref{eq:euclidean2})].}

Space $d\mathcal{S}$ is  conformally flat:  the above  coordinates can be used for it  (as ``stereographic'' coordinates), provided attention is given to the modified metric, 
\begin{equation}
g_{\mu \nu} =  \Omega^2(x) \, \eta_{\mu \nu}\, , \quad \mbox{with} \quad \,\,\, 
\Omega(x) = \frac{1}{1 - {\sigma^2}/{4 l^2}} \,\, .
\label{}
\end{equation}
$P_a$ and $K_a $ are, respectively, the generators of usual  translations (which define transitivity on  $\mathcal{M}$) and proper conformal transformations. Generators $L_{ab}$ refer to the Lorentz subgroup, whereas the remaining $L_{a4}$ define  transitivity on  the $d\mathcal{S}$  spacetime. 
Contact between de Sitter and Poncar\'e   groups is  made by  defining~\cite{gursey} the ``de Sitter translation generators''
\begin{equation}
\pi_a \equiv \frac{L_{a4}}{l} = P_a - l^{-2} K_a.\label{eq:dStransl}
\end{equation}
The Poincar\'e Lie algebra comes from the Lie algebra of the de Sitter group by the In\"on\"u--Wigner contraction $l \to \infty$~\cite{gursey,inonu,gil}.
 From the algebraic point of view, therefore, the change from Poincar\'e to de Sitter is achieved by replacing ordinary translation generators $P_a$ by the de Sitter ``translation'' generators $\pi_a$. The relative importance of   usual  translation and proper conformal generators in their contributions to $\pi_a$  is clearly determined by the value of $l$.  
 
The same contraction leads from a de Sitter spacetime to Minkowski space: $d\mathcal{S} \to \mathcal{M}$. And here another, major difference between $d\mathcal{S}$-relativity   and ordinary $\mathcal{M}$-based relativity emerges: while $\mathcal{M}$ is unique, $d\mathcal{S}$ is not. 
We have been talking about ``de Sitter group" and ``de Sitter space'', but it should be clear that there is an infinite number of   such groups and spaces~---~one for each value of the horizon $l$. 
 
A de Sitter spacetime shares with Minkowski spacetime a rare, remarkable  property: its Lorentzian bundle is itself a group: in the first case,  the de Sitter group~---~a principal bundle with $d\mathcal{S}$ as base space and the Lorentz group as fiber; in the second case, the bundle is  the Poincar\'e group~---~with $\mathcal{M}$ as base space and the Lorentz group again as typical  fiber.  An important result is that the Lorentz subgroup of  the de Sitter group preserves~\cite{lorentz} one very special length, just the horizon $l$. In consequence, the mentioned problem at the Planck scale is solved 
 provided, at ultra-high energies, the horizon $l$ identifies itself to the  Planck length $l_P$~---~which, by the way, illustrates  the necessary relationship between $l$ and the energy density involved.

Given  the Einstein equation
\begin{equation}
R^{\mu \nu } - {\textstyle\frac{1}{2}} R\ g^{\mu \nu }  =
 {\textstyle{ \frac{8\pi G}{c^{4}}  }} \, T^{\mu \nu} \ ,  \label{simpleEinstein}
\end{equation}%
 $\mathcal{M}$ is a solution with $T_{\mu \nu} = 0$. Space  $d\mathcal{S}$  can only be a solution with $T_{\mu \nu} = 0$ if an extra cosmological term $(-\, \Lambda g^{\mu \nu })$ is added to  the left-hand side, with forcibly $ \Lambda =  3 l^{-2}$. 
 In other words, Einstein's equation establishes a relationship between  the  pseudo--radius $l$ and  the cosmological constant $\Lambda$.   Minkowski space is a solution of 
\begin{equation}
 R^{\mu \nu } - {\textstyle\frac{1}{2}} R\ g^{\mu \nu } = 0\, ,
 \label{eq:Missolof}
\end{equation}%
 while de Sitter space is a solution of
\begin{equation}
 R^{\mu \nu } - {\textstyle\frac{1}{2}} R\ g^{\mu \nu } = \Lambda g^{\mu \nu } =  \frac{3}{l^2} \, g^{\mu \nu }.
\label{eq:dSissolof}
\end{equation}%
The standard procedure  encompasses both cases in the  Einstein equation  with a cosmological term,  
\begin{equation}
R^{\mu \nu } - {\textstyle\frac{1}{2}} R\ g^{\mu \nu }  - \Lambda g^{\mu \nu } =
 {\textstyle{ \frac{8\pi G}{c^{4}}  }} \, T^{\mu \nu} \ .  \label{cosmoEinstein}
\end{equation}%
The  left-- and the right--hand sides are separately  (covariantly) conserved quantitities. 

 The source in the right--hand side of (\ref{cosmoEinstein}) is the symmetric energy momentum tensor density. Recall that the passage $\mathcal{M} \to \mathcal{S}$ from Special to General Relativity is encapsulated in the minimal coupling prescription: the   Lorentz metric $\eta$ of  $\mathcal{M}$ is replaced by the Riemannian metric $g$ of
 $\mathcal{S}$,  and common derivatives are replaced by covariant derivatives defined by the Christofell-Levi-Civita metric connection of $g$.  In all these things  $\mathcal{M}$ must now be replaced by $d\mathcal{S}$. In particular,  translations have now to be considered on $d\mathcal{S}$, not  on $\mathcal{M}$. The source current is, in consequence, the  Noether current~\cite{coleman}
\begin{equation}
\Pi_{\mu \nu} =
T_{\mu \nu} - \frac{1}{l^2} \, {K}_{\mu \nu} 
\end{equation}%
for invariance under the  translations (\ref{eq:dStransl}) on $d\mathcal{S}$ including, besides  $T_{\mu \nu}$, the  symmetric version of the canonical proper conformal current 
\begin{equation}
K_{\mbox{\tiny{(c)}} \mu \nu}  =
\left(2 g_{\mu \lambda} \, x^\lambda x^\rho -
s^2 \delta_\mu^\rho\right) T_{\rho \nu} \, , \label{eq:confcurrent1}
\end{equation}%
where now $s^2 = g_{\mu \nu} x^{\mu} x^{\nu}$. Equation (\ref{cosmoEinstein}) must then be modified to 
\begin{equation}
R^{\mu \nu } - {\textstyle\frac{1}{2}} R\ g^{\mu \nu } - \Lambda g^{\mu \nu } =
 {\textstyle{ \frac{8\pi G}{c^{4}}  }} \, \left[ T_{\mu \nu} - \frac{1}{l^2} \, {K}_{\mu \nu}   \right] \ ,  \label{cosmoEinsteinmodif}
\end{equation}%
with translations on  $d\mathcal{S}$ replacing those on  $\mathcal{M}$.
Absence of sources ($T_{\mu \nu}  = 0$) has no more $\mathcal{M}$, but  $d\mathcal{S}$ for solution.  

Comparison with equations (\ref{eq:Missolof}) and  (\ref{eq:dSissolof}) suggests  particular relationships  between  pairs of terms: the  $\Lambda$-term and the $K$-term seem to have a  special link, as they both vanish when $l \to \infty$.  The $R$-terms, as the $T$-term, remain untouched in this limit. We shall prefer to emphasize these relationships by  rewriting  the general equation in the form
\begin{equation}
R^{\mu \nu } - {\textstyle\frac{1}{2}} R\ g^{\mu \nu }= R_{(T)}^{\mu \nu } - {\textstyle\frac{1}{2}} R_{(T)}\ g^{\mu \nu } - \Lambda g^{\mu \nu } =
 {\textstyle{ \frac{8\pi G}{c^{4}}  }} \, {\textstyle\left[ T_{\mu \nu} - \frac{1}{l^2} \, {K}_{\mu \nu}   \right]}.  \label{eq:generalEinsteinmodif}
\end{equation}
Only $R^{\mu \nu } - {\textstyle\frac{1}{2}} R\ g^{\mu \nu }$ and $T_{\mu \nu} - \frac{1}{l^2} \, {K}_{\mu \nu}  $ are conserved quantities. Neither $ R_{(T)}^{\mu \nu } - {\textstyle\frac{1}{2}} R_{(T)}\ g^{\mu \nu }$ nor $\Lambda g^{\mu \nu }$ are separately conserved. This means, in particular,  that $\Lambda $ can now be point-dependent. 

It has been stressed in previous presentations of de Sitter Special relativity~\cite{dssr,AP07a,lorentz,abap} that   the fundamental concepts of Special Relativity should be reexamined. Here, 
in a  first approach, we shall retain most of the  usual notions, such as those of energy density and pressure.  
Section \ref{sec:pocketSM} is an abstract of the Standard Cosmological Model with a cosmological constant.
Section \ref{sec:newparadigm} describes the main modifications caused by the change in local kinematics, and examine three consequential  issues:  (i)  the origin and value of the cosmological constant;  (ii) primeval inflation and (iii) the notion of comoving observers.

\section{Standard Model: a pocket version} 
\label{sec:pocketSM}
 The Standard Model  supposes for the large--scale spacetime $\mathcal{S}$ 
the Friedmann-Lema\^{\i}tre-Robertson-Walker (FLRW) interval~ \cite{FLRW}
\begin{equation}
ds^{2}= g_{\mu \nu} dx^{\mu} dx^{\nu}=c^{2} dt^{2}-a^{2}(t){\textstyle\left[\frac{dr^{2}}{1-kr^{2}}+r^{2} d\theta
^{2}+r^{2} \sin^{2}\theta d\phi ^{2}\right]}.  \label{zero1}
\end{equation}
It is convenient to take coordinate  $r$ dimensionless, as $\theta$ and $\phi$ ~---~the scale parameter $a(t)$ carrying   the length dimension.  The bracketed expression corresponds to the possible space sections  $\Gamma $: a sphere $S^3$ (if $k = + 1$), a hyperbolic hypersurface ${\mathcal H}^{2,1}$ (if $k = - 1$) or  an euclidean space ${\mathbb E}^3$ (if $k = 0$).  The energy content of the universe is modeled  by the energy-momentum of a perfect fluid~\cite{wein,kolb},
 \begin{equation} %
 T^{\mu\nu} = (p + \epsilon) \ U^{\mu} U^{\nu} - p \ g^{\mu\nu}  ,  
 \label{Tmunu}
 \end{equation} %
where $U^{\mu}$ is the four-velocity on a  streamline, 
$\epsilon = \rho c^2$ is the   energy density   and $p$ is the pressure. A fluid is perfect if it is isotropic when looked at from a (``comoving'')  frame carried  along its streamlines, that is, along the local integral curves of the four-velocities $U^{\mu}$.  These four-velocities submit to equations reflecting the conservation of $T^{\mu\nu} $ (see  Section \ref{sec:observers}). 

Interval (\ref{zero1}) incorporates 
two simplifying principles: the \textit{cosmological principle} and the
\textit{universal time principle}. 
The first postulates that the 
space-section $\Gamma $ is homogeneous and isotropic in the chosen coordinates.  The second principle states~\cite{nar, DiracFolh} that
the topology of the  four-dimensional manifold $\mathcal{S}$ is
the direct product\ $\mathbb{E}^1\times \Gamma$.  

Dynamics is governed by Einstein's equation  with a cosmological term (\ref{cosmoEinstein}).  This equation  determines, in principle, all acceptable spacetimes, or ``universes''. As already said, Minkowski space [with metric $\eta = $ diag $(1,-1,-1,-1) = \eta_{\alpha \beta} dx^{\alpha} dx^{\beta}$ in cartesian coordinates, equivalent to Eq.(\ref{zero1}) with $k = 0$ and $a(t) \equiv 1$] is a solution when $\Lambda $ and  $T^{\mu \nu}$ are zero. 

Once substituted into  Einstein's equation,  the metric  shown in interval (\ref{zero1})  leads to the Friedmann
equations for the scale factor $a\left( t\right) $:
\begin{equation} %
E(t) \equiv H^{2}+ \frac{k c^{2}}{a^{2}}= \frac{\dot{a}^{2}}{a^{2}} + \frac{k c^{2}}{a^{2}}=  \frac{\Lambda c^{2}}{3}+
\frac{8\pi G}{3 c^2} \, \epsilon~,  \label{Fried1}
\end{equation} %
\begin{equation} %
C(t) \equiv H^2+ {\dot H}  = \frac{\ddot{a}}{a} = \frac{\Lambda c^{2}}{3}-\frac{4\pi G}{3 c^{2}}\left( \epsilon
+  3 p  \right) ~.  \label{Fried2}
\end{equation} %
We have profited to recall the Hubble function $H(t)$ and to introduce  functions $E(t)$ and $C(t)$.  Let us further remember that present-day values are usually indicated by the index``0''. Today's value of the Hubble function, for example, is the Hubble constant $H_0$. And the red-shift parameter $z$ is defined by $1+z = \frac{a_0}{a(t)}$, in terms of  the scale parameter and its  present-day value $a_0$.  The energy density $\epsilon$ and pressure $p$ include those of usual matter, dark matter and radiation.   It will be convenient to use for them   the notations $\epsilon_m$ and  $p_m$  and introduce  the ``dark energy density''
\begin{equation}
\epsilon_{\Lambda} = \frac{\Lambda c^{4}}{8 \pi G}
\label{eq:darkenergy}\,\,\, . 
\end{equation} %
The cosmological term  is equivalent to a fluid with the ``exotic'' equation of state $\epsilon_{\Lambda} = -\, p_{\Lambda} =$ constant.   It is a  simplifying   trick to insert  $\epsilon_{\Lambda} = -\, p_{\Lambda} =  - {\textstyle{ \frac{1}{2}}}\left( \epsilon_\Lambda  + 3\, p_\Lambda \right) $ and write $\epsilon_{T},  p_{T}$ for the total energy and pressure. Equations (\ref{Fried1}, \ref{Fried2}) assume then the compact forms 
\begin{equation}
E(t) = {\textstyle{ \frac{8\pi G}{3 c^2}}} \left[ \epsilon_m +\epsilon_{\Lambda} \right] =  {\textstyle{ \frac{8\pi G}{3 c^2} }}\,  \epsilon_{T} ,   \label{Fried3} 
\end{equation} %
\begin{equation} %
C(t) =  {\textstyle{  \frac{8\pi G}{3 c^2}  }}\,  \left[ \epsilon_{\Lambda} - {\textstyle{ \frac{1}{2}}}\left( \epsilon_m  + 3\, p_m \right)  \right] =  -\,  {\textstyle{  \frac{4\pi G}{3 c^2}}}  \,  \left( \epsilon_T  + 3 p_T \right) .  \label{Fried4}
\end{equation} %
Function $E(t)$ is a measure of the  total  energy. Function $C(t)$ measures the scale parameter  concavity in terms of coordinate time. The middle expression in (\ref{Fried4}) exhibits clearly the contrasting effects of a (positive) cosmological constant and matter: $C < 0$ indicates slowing down expansion; $C > 0$, accelerated expansion. 

The  Christoffel symbols 
 of metric   (\ref{zero1}) can be all put  together  by using Kronecker deltas:\footnote{\, We assume   the summation convention   restricted to upper--lower contractions. Repeated lower--lower and upper-upper indices are not summed over. The usual relativistic $(0,1,2,3)$ notation will be extended to the indexing of matrix row and columns.}
\begin{equation*}
  \Gamma^{\alpha}{}_{\mu \nu}\, = \, 
 \delta_{\mu \nu} \Big\{\delta_{1 \mu} {\textstyle{\frac{\delta_0^{\alpha} a^2 H + \delta_1^{\alpha} k r}{1-k r^2}  }} \qquad \qquad   \qquad \qquad  
\end{equation*}
\begin{equation*}
 +  r \left[ \delta_0^{\alpha } a^2 H r  - \delta _1^{\alpha} \left(1-k r^2\right) \right]   (\delta _{2 \mu}+\delta_{3 \mu} \sin^2\theta) 
 - \delta_2^{\alpha} \delta_{3 \mu} \sin \theta  \cos \theta \Big\}
\end{equation*}
\begin{equation}
+ (\delta_{\lambda \mu} \delta^{\alpha}_{\nu} +  \delta_{\lambda \nu}  \delta^{\alpha}_{\mu})  (1-\delta_0^{\alpha}) 
\Big\{\delta_{0 \lambda} H + (\delta_2^{\alpha}+\delta_3^{\alpha})\delta_{1 \lambda}
{\textstyle{\frac{1}{r}  }}+ \delta_3^{\alpha} \delta_{2 \lambda}  \cot \theta     \Big\} .  \quad 
\label{christ1}
\end{equation}

 The Riemann tensor anti-symmetries in both the first and the second pairs of indices can be obscured by the metric factors necessary to raise and/or lower them. 
Specially simple are the   components with two upper and two lower indices,  $R^{\alpha \beta}{}_{\mu \nu}$.  Again by using  Kronecker deltas, we arrive at the  
expressions
\begin{equation*}
 R^{\alpha \beta}{}_{\mu \nu} =  {\textstyle{\frac{1}{c^2}}}\, \left(\delta_{\mu}^{\beta} \delta_{\nu}^{\alpha}-\delta_{\mu}^{\alpha } \delta_{\nu }^{\beta}  \right)
 \left[ C \left(\delta_{0 \nu} +  \delta_{0 \mu} \right) 
+ E  \left(1 - \delta_{0 \mu} -  \delta_{0 \nu} \right)    \right] \qquad  \quad
\end{equation*}
\begin{equation*}
\qquad \,\,\, =   {\textstyle{\frac{1}{c^2}}}\left(\delta_{\mu}^{\beta} \delta_{\nu}^{\alpha}-\delta_{\mu}^{\alpha } \delta_{\nu }^{\beta}  \right)
 \Big[ {\textstyle{ \left( H^2+ {\dot H}  \right)}} \left[ \delta_{0 \mu} + \delta_{0 \nu} \right]  
 \end{equation*}
\begin{equation}
\qquad \qquad  \qquad  \qquad + {\textstyle{\left( H^2+\frac{k  c^2}{a^2} \right)}}  \left(1 - \delta_{0 \mu} - \delta_{0 \nu}\right)\Big].
\label{astonish}
\end{equation}

The basic role of  the Einstein equation is to give the gravitational field (that is, the Riemann tensor) in terms of the source characters. These Riemann tensor components  depend on  the scale parameter and its derivatives exactly through the expressions which 
appear in the Friedmann equations and can, consequently, be written directly in terms of the source contributions:  Equations (\ref{Fried3}, \ref{Fried4}) lead immediately  to 
\begin{equation}       
R^{\alpha \beta}{}_{\mu \nu} =  {\textstyle{  \frac{8\pi G}{3  c^4}  }}  \left(\delta_{\mu}^{\beta} \delta_{\nu}^{\alpha}-\delta_{\mu}^{\alpha } \delta_{\nu }^{\beta}  \right)
 \left\{ \epsilon_{T}  - {\textstyle{ \frac{3}{2} }} ( \epsilon_T  +  p_T)  \left( \delta_{0 \nu} +  \delta_{0 \mu} \right)  \right\}  \, .
\label{astonish2}
\end{equation}
Such components depend, consequently,  only on the  time coordinate. 
The Ricci tensor,  given by
\begin{eqnarray}\nonumber
 R^{\alpha}{}_{\mu}  =  -\,  {\textstyle{\frac{1}{c^2}}}\,  \delta^{\alpha}_{\mu} \left[ C (1+2 \delta_{0 \mu}) + 2 E  (1-  \delta_{0 \mu})  \right]  \qquad  \qquad  \qquad  \qquad\\
\quad  \quad \, = -\,  {\textstyle{\frac{1}{c^2}}}\,  \delta^{\alpha}_{\mu} \left[ {\textstyle{ \left( H^2+ {\dot H}  \right)  }}  (1+2 \delta_{0 \mu}) + 2 {\textstyle{  \left( H^2+\frac{k}{a^2} \right) }}  (1-  \delta_{0 \mu})  \right] ,
\label{Riccitensor0}
\end{eqnarray}
will constitute  the matrix
\begin{equation*}
(R^{\alpha}{}_{\mu})  =   -\, {\textstyle{ \frac{1}{c^2} }} {\small{ \left(
\begin{array}{cccc}
3 C & 0 & 0 & 0 \\
 0 & 2 E  + C & 0 & 0 \\
 0 & 0 &  2 E  + C & 0 \\
 0 & 0 & 0 & 2 E  + C 
\end{array}
\right) }}
\end{equation*}
\begin{equation}
\qquad \qquad=  {\textstyle{ \frac{4 \pi G}{ c^4} }} {\small{\left(
\begin{array}{cccc}
\epsilon_T + 3 p_T & 0 & 0 & 0 \\
 0 & p_T - \epsilon_T& 0 & 0 \\
 0 & 0 & p_T - \epsilon_T & 0 \\
 0 & 0 & 0 & p_T - \epsilon_T
\end{array}
\right) }}, 
\label{Riccitensor}
\end{equation}
whose trace is the scalar curvature
\begin{equation}
 R = -\, {\textstyle{ \frac{6}{c^2} }} (E  + C) = -\, {\textstyle{  \frac{6}{c^2} }}\left[ 2 H^2 + {\dot H} + \frac{k c^2}{a^2}  \right] ={\textstyle{  \frac{ 8 \pi G}{ c^4}}}\, (3\, p_{T} -  \epsilon_{T})\,\, .
\label{Rscalar}
\end{equation}

From this equation comes a  general ``equation of state'', which  must  hold for any homogeneous isotropic model: 
\begin{equation}
p_{T} = \frac{ \epsilon_{T}}{3} +\frac{R\,  c^4}{24 \pi G}\, \, .
\label{eq:generalEOS}
\end{equation}

The general  Einstein tensor, finally, will be 
\begin{equation*}
(G^{\alpha}{}_{\mu}) = (R^{\alpha}{}_{\mu} - {\textstyle{\frac{1}{2}}} \delta^{\alpha}_{\mu} R)  = 
\left(
\begin{array}{cccc}
\frac{3 E}{c^2} & 0 & 0 & 0 \\
 0 &  \frac{2 C+E}{c^2}   & 0 & 0 \\
 0 & 0 &  \frac{2 C+E}{c^2}   & 0 \\
 0 & 0 & 0 &  \frac{2 C+E}{c^2} 
\end{array}
\right) 
\end{equation*}
\begin{equation}
\qquad = 
{\textstyle{ \frac{8\pi G}{c^4} }} \, \left(
\begin{array}{cccc}
  \epsilon_{T}  & 0 & 0 & 0 \\
 0 & - \, p_{T}  & 0 & 0 \\
 0 & 0 &  - \, p_{T}   & 0 \\
 0 & 0 & 0 & - \, p_{T}  
\end{array}
\right). \label{Etensor} 
\end{equation}
This  is  simply an alternative version of  Einstein's equation (\ref{cosmoEinstein}).

The Friedmann equations acquire their most convenient form
in terms of dimensionless variables, obtained by dividing them by  the squared Hubble constant: define the parameters
\begin{equation}
\rho _{crit}=\frac{3H_{0}^{2}}{8\pi G} \, ;\,  \Omega
_{m} = \frac{\rho_{m}}{\rho _{crit}} =  \frac{8\pi G \rho_{m}}{3 H_{0}^{2}}  \, ;\, \, 
\end{equation}
\begin{equation}
\Omega_{\Lambda} = \frac{\Lambda c^{2}}{3H_{0}^{2}} =
\frac{\rho_{\Lambda}}{\rho_{crit}}  \, ;\, \,  \Omega_{{\kappa}}(t) = - 
\;\frac{{\kappa}c^{2}}{a^{2}H_{0}^{2}}\, ;\,\,   \omega = \frac{p_m}{\epsilon_m}.
\label{eq:Omegas}
\end{equation}
Parameter $\omega$ is a frequently used convenience, here taken in its most general form: it can be $\epsilon$-dependent. 
Then,
\begin{equation}
 \frac{E}{H_0^2} - \frac{\kappa c^2}{a^2 H_0^2} = \frac{H^2}{H_0^2} =  \Omega_{\Lambda } +\Omega_{{\kappa}}+ \Omega_{m}\, ; \label{eq:ECfrmOms1}
\,\,
\frac{C}{H_0^2} =  \Omega_{\Lambda } - \frac{1+3\omega}{2} \,\, \Omega_{m}. 
\end{equation}
Equation (\ref{eq:ECfrmOms1}) leads, for present-day values, to the usual normalization of observed parameters, $ \Omega_{\Lambda } + \Omega_{{\kappa 0}}+ \Omega_{m0} = 1$. Observations  (crossed supernova/BBR data) favor the values $ \Omega_{{\kappa}} \approx  0$,  $\Omega_{\Lambda }  \approx 3/4$ and $\Omega_{m0}  \approx  1/4$, the latter including all visible matter, radiation and dark matter.

Notation is still  further simplified, and expressions put into closer contact with these measured quantities,   if we  define   new  objects as:
\begin{equation} \nonumber 
{\mathcal R}^{\alpha \beta}{}_{\mu \nu} = \frac{c^2}{ H_{0}^{2}} \, R^{\alpha \beta}{}_{\mu \nu} , \,\,\,
{\mathcal R^{\alpha}{}_{\mu}} =  \frac{c^2}{ H_{0}^{2}} \, R^{\alpha}{}_{\mu} , \,\,\,  {\mathcal R} = \frac{c^2}{H_{0}^{2}} \,R  \, , 
\end{equation}
\begin{equation}
{\mathcal G^{\alpha}{}_{\mu}}  =  \frac{c^2}{ H_{0}^{2}} \,G^{\alpha}{}_{\mu}\,  \,\,\, {\mbox{and}} \,\, \,\,\,{\mathcal T^{\alpha}{}_{\mu}}  =  \frac{8 \pi G }{c^2 H_{0}^{2}} \,T^{\alpha}{}_{\mu}\, .
\label{eq:caltensors}
\end{equation}
Equation (\ref{Etensor} ) becomes 
\begin{equation*}
({\mathcal G^{\alpha}{}_{\mu}}) = ({\mathcal R^{\alpha}{}_{\mu}}  - {\textstyle{\frac{1}{2}}} \delta^{\alpha}_{\mu} {\mathcal R} )  
= ({\mathcal T^{\alpha}{}_{\mu}})
\end{equation*}
\begin{equation}
 \quad = 3
{\textstyle{ \left(
\begin{array}{cccc}
 \Omega_{\Lambda }  + \Omega_{m}\ & 0 & 0 & 0 \\
 0 &  \Omega_{\Lambda }  - \omega\,  \Omega_{m}  & 0 & 0 \\
 0 & 0 &  \Omega_{\Lambda }  - \omega\,  \Omega_{m}   & 0 \\
 0 & 0 & 0 &  \Omega_{\Lambda }  - \omega\,  \Omega_{m} 
\end{array}
\right)  }}. \label{Ecaltensor} 
\end{equation}
Taking traces give now 
\begin{equation}
{\mathcal  R} + 12\,   \Omega_{\Lambda } =  3\,   (3 \omega - 1)\,  \Omega_{m} =  -\,\, {\mathcal T}_m. 
\label{eq:traces}
\end{equation}
We have transferred back the $\Lambda$ contribution to the left hand side, leaving on the right only the trace of the matter energy-momentum. 

%
\section{The new paradigm}
\label{sec:newparadigm}
As a first step to obtain the modified Einstein equation,  
we move the dark-energy term (\ref{eq:darkenergy})   from the right-hand side of Eq. (\ref{Etensor})  back  to the left-hand side, getting
\begin{equation}
(G^{\mu }{}_{\nu }-\Lambda  \delta_{\nu }^{\mu}) = {\textstyle{ \frac{8  \pi G }{c^4} }} \, 
\left(
\begin{array}{cccc}
 \epsilon_m & 0 & 0 & 0
\\
 0 &- p_m &  0 & 0
\\
 0 & 0 & - p_m & 0
\\
0 & 0 &  0 & - p_m
\end{array}
\right)\, .
\label{eq:firststep}
\end{equation}
Use of Eqs.(\ref{eq:caltensors}), and of  $p_m = \omega\,  \epsilon_m$ for the equation of state   leads then to
\begin{equation}
({\mathcal G}^{\mu }{}_{\nu }- 3\, \Omega_{\Lambda}  \delta_{\nu }^{\mu})  = \, 3\,   \Omega_m \, 
\left(
\begin{array}{cccc}
1 & 0 & 0 & 0
\\
 0 &- \omega &  0 & 0
\\
 0 & 0 & - \omega & 0
\\
0 & 0 &  0 & -\omega
\end{array}
\right)\, .
\label{eq:firststep2}
\end{equation}

The conformal source term must now be added in the  right-hand side.  We arrive at the new fundamental equation
\begin{equation}
R^{\mu }{}_{\nu } - {\textstyle{ \frac{1}{2} }} \,  R \, \delta _{\nu }^{\mu }= R_{(T)}^{\mu }{}_{\nu } - {\textstyle{ \frac{1}{2} }} \,  R_{(T)} \, \delta _{\nu }^{\mu } -\Lambda\,   \delta_{\nu }^{\mu }=  {\textstyle{ \frac{8  \pi G }{c^4} }} \, 
\Pi^{\mu}{}_{\nu} \, ,
\label{eq:newE}
\end{equation}
where 
\begin{equation}
\Pi^{\mu}{}_{\nu} = T^{\mu}{}_{\nu} - {\textstyle{\frac{1}{l^2} }}\,  K^{\mu }{}_{\nu} 
\label{eq:Pi}
\end{equation}
is the Noether current for invariance under translations on a de Sitter spacetime, modified by the minimal coupling prescription.  Tensor $R^{\mu }{}_{\nu } $ is the true spacetime Ricci tensor. Symbol $R_{(T)}^{\mu }{}_{\nu } $ is an artifact,   tentatively separating the energy-momentum contribution from that of the conformal current.

The conformal current (\ref{eq:confcurrent1})  will be, in stereographic coordinates, 
\begin{equation}
K^{\mu}{}_{\nu} = x^{\lambda} \left[x^{\mu} T_{\lambda \nu} + x_{\nu} T_{\lambda \mu} \right] - s^2 T^{\mu}{}_{\nu}. 
\label{eq:concurr}
\end{equation}
It is again convenient to introduce, in a way analogous to Eqs.(\ref{eq:caltensors}), the modified current
\begin{equation} \nonumber 
{\mathcal K}^{\mu}{}_{\nu}  = \frac{c^2}{ H_{0}^{2}} \, K^{\mu}{}_{\nu} = 
 x^{\lambda} \left[x^{\mu} {\mathcal T}_{\lambda \nu} + x_{\nu} {\mathcal T}_{\lambda \mu} \right] - s^2 {\mathcal T}^{\mu}{}_{\nu}, 
\label{eq:calK}
\end{equation}
as well as its modified trace ${\mathcal K}$. We shall only need traces afterwards, and shall  move to  convenient coordinates. 
In  comoving FLRW coordinates  for $\kappa = 0$, interval (\ref{zero1}) reduces to 
\begin{equation}
ds^{2} =   c^{2} dt^{2}-a^{2}(t)\left[dx^{2} +dy^{2} +dz^{2} \right] \,  .  \label{zerok=0}
\end{equation}
The trace  is then 
\begin{equation}
{\mathcal K} = 3\, \Omega_m \left[ (1 - \omega) \, s_E^2   + 4 \omega\, c^{2} t^{2}\right]
\label{eq:traceofK}
\end{equation}
where the euclideanized, positive version of the squared finite interval 
\begin{equation}
s_E^2=  c^{2} t^{2} + a^{2}(t)\left[x^{2} + y^{2} +z^{2} \right]  \ge 0 
\label{eq:euclidean2}
\end{equation}
makes its appearance.  We see that, for physically reasonable values of $\omega$,  trace ${\mathcal K}$  is  always positive.

Taking the trace of (\ref{eq:newE})  leads to [now with notation $d \equiv a(t) r$]
\begin{equation*}
\Lambda + \frac{R_{(T)}}{4} = - {\textstyle{ \frac{2  \pi G }{c^4} }}   \left\{{\textstyle{\left(1+ \frac{c^2 t^2-d^2}{l^2} \right)}} \left(\epsilon_m-3 p_m\right) -
{\textstyle{\frac{2 d^2}{l^2} }}p_m - {\textstyle{\frac{2 c^2 t^2}{l^2} }}\epsilon_m
\right\}  .
\end{equation*}

Using (\ref{eq:traceofK}), the modified form  of Eq.(\ref{eq:traces}) in terms of the $\Omega$'s  will be
\begin{equation}
{\textstyle{\frac{1}{12}}} {\mathcal R}_{(T)} + \Omega_ \Lambda = {\textstyle{\frac{1}{4}}} \, \Omega_{m} \Big[ \frac{(1 -  \omega) \, s^2_E + 4\, \omega\,  c^2 t^2}{l^2} - (1 - 3 \omega)
\Big] 
\, .
\label{eq:tracesmodifOm}
\end{equation}
The last term inside  the bracket is the energy-momentum contribution, while the conformal current contribution is tagged by the $1/l^2$ factor. Notice a consonance of signs:  for reasonable values of \,\,\,$ \omega$\,\, the   ``curvature'' $R_{(T)} $ is, as the 
 energy-momentum contribution, negative; the $\Lambda$ term is positive,  as is  the conformal current contribution. 

\subsection{The cosmological constant}
\label{subsec:CC}

To obtain an order-of-magnitude estimate of the present-day value of $\Lambda$   we are tempted to equal ${\mathcal R}_{(T)} $  to the usual, $T$ -- driven term in (\ref{eq:tracesmodifOm}): 
\begin{equation}
{\mathcal R}_{(T)}   = -\,  3\,   \Omega_{m}   (1 - 3 \omega) , 
\label{eq:Tterms}
\end{equation}
and write
\begin{equation}
\Omega_ \Lambda = {\textstyle{\frac{1}{4}}}\,  \Omega_{m} \Big[ \frac{(1 -  \omega) \, s^2_E + 4\, \omega\,  c^2 t^2}{l^2}  \Big]  . 
\label{eq:lambdamaybe}
\end{equation}
As at present dust ($p_m << \epsilon _m$, or $\omega \approx 0$) dominates,
\begin{equation}
\Omega_ \Lambda = {\textstyle{\frac{1}{4}}}\, \Omega_{m}  \,\, \frac{ s^2_E }{l^2} \,\,  . 
\label{eq:Omlambdamaybe}
\end{equation}

Let us, then, proceed to a closer examination of the euclideanized  interval (\ref{eq:euclidean2}). For any realistic calculation,\footnote{\ For all experimental/observational data we shall be using references  [\cite{WMAP}] and [\cite{ParticleData}]. }  its expression
\begin{equation} \nonumber
s_E^2 =  c^{2} t^{2} + a^{2}(t) \,r^{2}  \, 
\end{equation}
presents a clear problem: $a(t)$ is obtained, in each case, by an integration and includes an arbitrary integration constant.
Term $c^{2} t^{2} $  does not.  Present-day value of $c t$ can be taken as  $c t \approx \frac{c}{H_0} \approx 9.26 \times 10^{27} h^{-1}$ cm $\approx 13.2 \times 10^{27} $ cm, corresponding to $t \approx 13.8 \times 10^{9}$ years.  

To fix the integration constant in $a(t)$ we should know the "size" of the Universe at some value of  time. Let us then obtain  a particular rough  estimate which has the advantage of letting clear where eventual errors can lie in wait.  Take the recombination time as $t_r  \approx 3.8 \times 10^5$  years $ \approx  1.19 \times 10^{13}$ sec. And suppose (first possible error, see below) the size at that time to be $d_r \approx  c t_r  \approx  3.58 \times  10^{23}$  cm. The present day value $d_0$ would then be given by a simple multiplication by the red-shif $1 + z \approx 1100$: $d_0 \approx  3.94 \times  10^{27}$ cm. This would be the Universe size, that is, the size of the domain contributing to the cosmological constant: 
$d_0 = l$. If that is so, then $ct \approx \frac{c}{H_0} \approx \frac{13.2}{3.94}\,  l  \approx 3.36\, l$. This would give
\begin{equation}
s_E^2 = c^{2} t^{2} + l^{2} \approx   ( 3.36)^2 l^{2} + l^{2}\approx  12.3 \,  l^2. 
\label{eq:sEestimate}
\end{equation}
Equation (\ref{eq:Omlambdamaybe}) would then give the value
\begin{equation}
\Omega_ \Lambda \approx  3.07 \,\,  \Omega_{m} ,
\label{eq:OmegLambvalue}
\end{equation}
very close to that given by observations.  Of course, $\Omega_{{\kappa 0}} = 0$  would then lead to $\Omega_m \approx 1/4$.

A comment on the possible sources of error. (i) Assumption (\ref{eq:Tterms}) is, of course, rather arbitrary; nevertheless, the argument could be reversed: insertion of the observation values in the above formulae would lead to that condition;  (ii) the value of  $d_r$ seems overestimated: it has been taken as the distance free light would traverse up to  recombination time; notice, however, that  (iii)  inflation has not been considered, which could enlarge initial lengths anyhow;   (iv)  passage from radiation-- to dust--dominated eras is a continuous process; 
a more reliable value could come from numerical calculations, as the complete solution can only be written in terms of rather involved elliptic functions~\cite{ACM05}.

The important point is that, in the new paradigm, $\Omega_ \Lambda$ is no more a free parameter: it has a definite physical meaning as an effect of the conformal current, and is determined~---~like that same current~---~by the matter content.

\subsection{Primeval inflation}
\label{subsec:inflation}

A first approach to the problem of primordial inflation can be made under certain rough assumptions.  One is that the primeval Universe is dominated by radiation, and that interactions (for instance, of particles produced by pair production) can be neglected so as to keep valid the equation of state $p_\gamma = \epsilon_\gamma/3$, or $\omega = 1/3$. Within this model, Eq.(\ref{eq:tracesmodifOm}) becomes 
\begin{equation}
{\textstyle{\frac{1}{12}}} {\mathcal R}_{(T)} + \Omega_ \Lambda = {\textstyle{\frac{1}{6}}} \Omega_{\gamma}  \frac{s^2_E + 2 c^2 t^2}{l^2} = {\textstyle{\frac{1}{6}}} \Omega_{\gamma}  \frac{a^{2}(t) r^{2}  + 3 c^2 t^2}{l^2}  \, .
\label{eq:forinfl1}
\end{equation}
A second hypothesis is that radiation, which by  Eq.(\ref{eq:generalEOS}) does not contribute to the usual curvature, here does not contribute to ${\mathcal R}_{(T)}$, which consequently vanishes.  Function $\Omega_{\gamma}$, the part of $\Omega_{m}$ stemming from radiation,   is given by 
\begin{equation}
\Omega_{\gamma} = \Omega_{\gamma 0}(1+z)^4 = 2.46 \times 10^{-5} h^{-2}  (1+z)^4 \approx  5. \times 10^{-5}  (1+z)^4. 
\label{eq:omgamma}
\end{equation}

On the other hand, in a radiation--dominated Universe,  $\frac{a (t)}{a_0}= \frac{0.082 \sqrt{H_0}}{h^2} \sqrt{t} $, so that $\frac{a^2 (t)}{a^2_0} \approx 6.1 \times 10^{-20} t $\, and $(1+z) \approx \frac{4. \times 10^{9}}{\sqrt{t}}$. The term in $c^2 t^2$ in
\begin{equation}
 \Omega_ \Lambda = {\textstyle{\frac{1}{6}}} \Omega_{\gamma}  \frac{a^{2}(t) r^{2}  + 3 c^2 t^2}{l^2}  
\label{eq:forinfl2}
\end{equation}
is then easily obtained: $ 3 c^2 t^2 \approx \frac{7.3 \times 10^{59}}{(1+z)^4}$.  The term $a^{2}(t) r^{2} $ presents the same difficulty discussed in the previous paragraph, and we resort to an analogous approach: we take $a_0 = d_r = c t_r$, so that $a^{2}(t) r^{2} \approx \frac{3.6  \times 10^{23}}{(1+z)^2}$. Putting again $a(t) r  = l$, we arrive at 
\begin{equation*}
 \Omega_ \Lambda \approx  0.83 \times 10^{-5} (1+z)^4 \left(1+ \frac{2 \times 10^{36}}{(1+z)^2}    \right)  
\end{equation*}
\begin{equation}
\qquad \,\,\, \approx   
 2.13 \times 10^{33}  \left(\frac{1}{t^2} + \frac{1.25  \times 10^{17}}{t}    \right) .
\label{eq:inflformula}
\end{equation}

Equality between the two  terms takes place at $t  \approx  8. \times 10^{-18} $ sec. 
As $\Omega_{\Lambda} = \frac{ \Lambda c^2}{3 H_0^2}$, 
\begin{equation}
\Lambda \approx   
3.8  \times 10^{-21}  \left(\frac{1}{t^2} + \frac{1.25  \times 10^{17}}{t}    \right)  cm^{-2}.
\label{eq:inflformula}
\end{equation}
We have thus a highly accelerated expansion for very small values of $t$. 

\subsection{Comoving observers}
\label{sec:observers}
In standard General Relativity, comoving observers are fixed by 
energy-mo\-men\-tum  conservation, which  requires the vanishing of  the covariant divergence of  (\ref{Tmunu}). That  leads  to~\cite{ellis,livro}
\begin{equation}
 (p + \epsilon)\, \nabla_U U^{\mu}  = (g^{\mu \nu} -  U^{\mu} U^{\nu})\, \partial_{\nu} p\, .
\label{notgeods}
\end{equation}
This means that the observer going along the streamlines is, in the general case, accelerated. Expression
\begin{equation}
P^{\mu \nu} = g^{\mu \nu} -  U^{\mu} U^{\nu}
\label{eq:Uprojector}
\end{equation}
is a projector~--~what we have in the right-hand side is the projection, orthogonal to the four-velocity  $U$, of the pressure spacetime gradient.  Space homogeneity imposes $\partial_{k} p = 0$, so that actually
\begin{equation}
 (p + \epsilon)\,  \nabla_U U^{\mu}  = (g^{\mu 0} -  U^{\mu} U^{0})\, \partial_{0} p\, .
\label{notgeods2}
\end{equation}
The flux lines will be geodesics only for constant $p$. It is the case of  a dust fluid, and of a dust fluid plus standard dark energy.  For pure dark energy, no velocities appear in  $T_{(\Lambda)}^{\mu \nu} = \epsilon_{\Lambda}\, g^{\mu \nu} $, which is automatically preserved. In this  case Eq.(\ref{notgeods}) gives no information at all: both of its sides vanish. 

 The matrix in Eq.(\ref{Etensor}) is just the energy-momentum of an isotropic fluid, as seen by an  observer attached to a frame going along its flux lines with four-velocity  $U$.  FLRW coordinates are appropriate just  for  such observers, which will see the metric in the form of Eq.(\ref{zero1}).

In the new paradigm, the comoving observers are no more those given by Eqs.(\ref{notgeods}) or (\ref{notgeods2}). They must now be consistent with Eq.(\ref{eq:newE}), whose right-hand side must have vanishing 
 covariant divergence. The result is 
\begin{eqnarray} \nonumber
\left[1 + \frac{s^2}{l^2} \right] (\epsilon + p)  \triangledown _U U_{\mu} 
- P _{\mu }^{\sigma} \partial_\sigma p \qquad \qquad \qquad  \qquad \qquad   \qquad \qquad \qquad 
\\  \nonumber
=  \frac{1}{l^2}  P _{\mu }^{\sigma}  \Big[p  \triangledown _\sigma  (s^2)  
+  \triangledown^\nu \left\{  2 \epsilon  x_{\nu } x_{\sigma }   - (p+\epsilon) x_{\rho} x_{\lambda} 
P _{\beta }^{\rho } ( \delta _{\nu }^{\beta } \delta _{\sigma }^{\lambda }+\delta _{\sigma }^{\beta } \delta _{\nu }^{\lambda })
  \right\} \Big] , 
\label{eq:observer1}
\end{eqnarray}
which is the same as
\begin{eqnarray}  \nonumber
(p+\epsilon )  \triangledown _U U_{\mu} 
-  P _{\mu }^{\sigma} \partial_\sigma p = 
 -\,  \frac{1}{l^2}  \Big[(\epsilon + p) s^2  \triangledown _U U_{\mu}  
  \qquad \qquad   \qquad \qquad \quad \\
\qquad-  P _{\mu }^{\sigma} 
\triangledown^\nu \left\{  \left[2 \epsilon \delta _{\sigma }^{\lambda } \delta _{\nu }^{\rho} - 
(\epsilon + p) P_{\beta }^{\rho } ( \delta _{\nu }^{\beta } \delta _{\sigma }^{\lambda }+\delta _{\sigma }^{\beta } \delta _{\nu }^{\lambda }) \right] x_{\rho} x_{\lambda} 
  \right\}
\Big]\, .
\label{eq:observer2}
\end{eqnarray}

We have made repeated use of projector (\ref{eq:Uprojector}). This expression maintains the  orthogonal velocity--acceleration property, and reduces to Eq.(\ref{notgeods}) when $l \to \infty$.  In the new paradigm, FLRW coordinates are appropriate just  for   observers satisfying Eq.(\ref{eq:observer2}). \\

\section{Final  comments}
\label{sec:finam}

Physical Cosmology is perhaps the boldest enterprise of Physical Science~---~trying to describe the far out Universe with the  laws it has found around the Earth. It has only recently become a phenomenology, and it is not surprising that some of its results are baffling.  The very presence of a cosmological constant is bewildering enough.  
It is surely   desirable that  an explanation for it  be found  within the scheme of local Physics, however modified. 
We have seen that this may be the case, provided local kinematics is described by de Sitter Relativity instead of ordinary Poincar\'e Relativity.

Progress in Physics has been  achieved through  the use of a succession of correspondence principles.  Special Relativity has for General Relativity the role classical physics has for quantum physics. Any change in the first will find its echos in the second. 
The local kinematics of Special Relativity underpins that of General Relativity. 
 Changing from Poincar\'e to de Sitter local kinematics impinges heavily on the more general kinematics. 

We think of General Relativity as the dynamics of the gravitational field. In this sense, we can look at the presence of local ``cosmological constants'' as a dynamical effect.  Actually, gravitation is the only interaction which complies to Hertz's program of reducing dynamics to kinematics.  It  changes spacetime itself, it redefines its kinematics.  The other fundamental interactions exhibit no such character~---~they are, in this sense, more essentially dynamical. On the other hand, they are described within the gauge paradigm, which is more receptive to conformal symmetry. Einstein's sourceless equation is not conformally invariant, while the Yang-Mills equation is. It may well be that the passage from Poincar\'e to de Sitter relativities~---~encapsulated in the addition of the proper conformal transformations to usual translations~---~have far less traumatic consequences for electromagnetic, strong and weak interactions than it has for gravitation.

\section*{Acknowledgments}
The authors would like to thank J. P. Beltr\'an Almeida and C. S. O. Mayor for useful discussions. They would like to thank also FAPESP, CNPq and CAPES for partial financial support.



\end{document}